\documentclass[12pt,a4paper,dvips]{article}

\usepackage{epsfig}
\usepackage{graphicx}
\usepackage{array}
\usepackage{dcolumn}
\usepackage{multicol}
\usepackage{cite}
\mathchardef\d="0064            
\def\diameter{\kern0.3em/\kern-0.72em$\bigcirc$}
\begin{document}

\begin{center}\textbf{\Large {\bf LFS-3 - new radiation hard 
scintillator for electromagnetic calorimeters.}}\end{center}

\vspace{1.cm}

\begin{center}
{\large 
V.A. Kozlov$^{1}$,  
A.I. Zagumennyi$^{2}$, 
Yu.D. Zavartsev$^{2}$, \\
M.V. Zavertyaev$^{1}$,
A.F. Zerrouk$^{3}$}
\end{center}

\vspace{0.5cm}

{\normalsize 1. P.N. Lebedev
Physical Institute of Russian Academy of Sciences, Moscow, Russia} 
\vspace*{0.5cm}

{\normalsize 2. Prohorov General Physics Institute of Russian Academy of Sciences, 
Moscow, Russia}\vspace*{0.5cm}

{\normalsize 3. Zecotek Imaging Systems Pte Ltd, Division 
of Zecotek Photonics Inc., Vancouver, Canada}

\vspace{1.5cm}

\Large {\bf Abstract}

\bigskip

\normalsize
 Radiation damage of new heavy $LFS-3$ scintillating crystals has been
studied using powerful $^{60}Co$ source at the dose rate of 4 Krad/min. No 
deterioration in optical transmission of $LFS-3$ crystals was observed after irradiation 
with the dose of 23 Mrad.

\section{Introduction}

 In the last years extensive effort has been directed to develop
new scintillating materials for Positron Emission Tomography (PET).
These materials must be characterized by high light yield, a fast
scintillation decay time, a good energy resolution and small absorption
length for gamma radiation.

 Fast and dense scintillating crystals $Lu_{2}SiO_{5}(LSO)$ were discovered 
and investigated by C.L. Melcher and J.S.Schweitzer in 1992 as promising 
material for gamma-ray detection \cite{Melcher1,Melcher2}. 
  Later lutetium-yttrium oxyorthosilicate $(LYSO)$
crystals were discovered and tested \cite{Cooke}.

Stoichiometric lutetium oxyorthosilicate $(Ce_{x}Lu_{1-x}SiO_{5}$, LSO) and 
yttrium substituted $LYSO$ $(Ce_{x}(Lu,Y)_{1-x}SiO_{5})$ crystals are currently 
commercially grown for the fabrication
of scintillation detectors. The large-size $LSO/LYSO$ crystals were proposed for 
application in future high-energy physics experiments as attractive materials 
for homogeneous high resolution electromagnetic (EM) calorimeters 
\cite{Chen1,Chen2}. Recently first 
prototype of EM calorimeter based on $3\times3$ large volume $LYSO$ crystals was 
successfully tested at MAMI accelerator with photons up to 490 MeV energy 
\cite{Thiel} .

 The one drawback of $LSO$ is the relatively large spread of its scintillation parameters 
within the boule (top and bottom) and between different boules. The advertised $LYSO$ crystals 
show slightly better light yield efficiency and decay time as compared to $LSO$. They, however, 
exhibit similar non-uniformity of scintillation parameters across the boule, with an additional 
intrinsic tendency to cracking.

\section{Results and discussion}

\begin{table}[htp]
\caption{The basic properties of the scintillating crystals.}
\label{tab:prop}
\begin{center}
\begin{tabular}{|c|c|c|} \hline
\normalsize Material & \normalsize $NaI(Tl)$ 
&\normalsize LFS-3 
\\ \hline\hline
\normalsize Density, $\rho$ $(g/cm^{3})$   &\normalsize   3.67 
&\normalsize   7.35 
\\ \hline

\normalsize Melting point, $(^{0}C)$       &\normalsize   651  
&\normalsize   2000 
\\\hline

\normalsize Radiation length, $X_{0}$ (cm) &\normalsize   2.59 
&\normalsize   1.15 
\\\hline

\normalsize Moliere radius, $R_{m}$ (cm)   &\normalsize   4.3  
& \normalsize  2.09 
\\\hline

\normalsize Light output ($\%$)            &\normalsize   100  
& \normalsize   85  
\\\hline

\normalsize Decay time, (ns)               &\normalsize   230  
&\normalsize    35  
\\\hline

\normalsize Peak emission, (nm)            &\normalsize   410  
& \normalsize  425  
\\\hline

\normalsize Refractive index, $n$          &\normalsize   1.85 
& \normalsize  1.81 
\\
\normalsize in maximum of emission         &              
&              
\\\hline

\normalsize Hardness, (Moh)                &\normalsize    2   
&\normalsize    5   
\\\hline

\normalsize Hygroscopic                    &\normalsize   Yes  
&\normalsize   No   
\\\hline

\end{tabular} 
\end{center}
\end{table}

Proprietary, bright scintillators $LFS-3$ (Lutetium Fine Silicate) developed by 
Zecotek Imaging Systems Pte Ltd provide much improved scintillating parameters and 
\mbox{reproducibility \cite{Zagumennyi}}. 
$LFS$ is a brand name of the set of Ce-doped scintillation crystals of the solid 
solutions on the basis of the silicate crystal, comprising lutetium and
crystallizing in the 
monoclinic system, spatial group $C2/c$, $Z=4$. The patented $LFS-3$
compositions is $Ce_{x}Lu_{2+2y-x-z}
A_{z}Si_{1-y}O_{5+y}$, where A is at least one 
element selected from 
the group consisting of $Ca$, $Gd$, $Sc$, $Y$, $La$, $Eu$ and $Tb$. 

%

The raw materials were $99.999\%$ pure $Lu_{2}O_{3}$, $SiO_{2}$ and the scintillating $CeO_{2}$ dopant. The 
$LFS$ crystals demonstrated stable scintillation parameters for top and bottom of large boules in 
comparison with $LSO$. The most important parameters of $LFS$ scintillating crystals are presented 
in Table\,\ref{tab:prop} in comparison with characteristics of common inorganic
scintillator $NaI(Tl)$. The main 
properties of $LFS$ crystal make it highly suitable as a scintillating material for electromagnetic 
calorimeters in high energy particle physics experiments.

\begin{figure}[htb] 
\addtolength{\abovecaptionskip}{6pt}
\centering
\includegraphics[width=8.5cm]{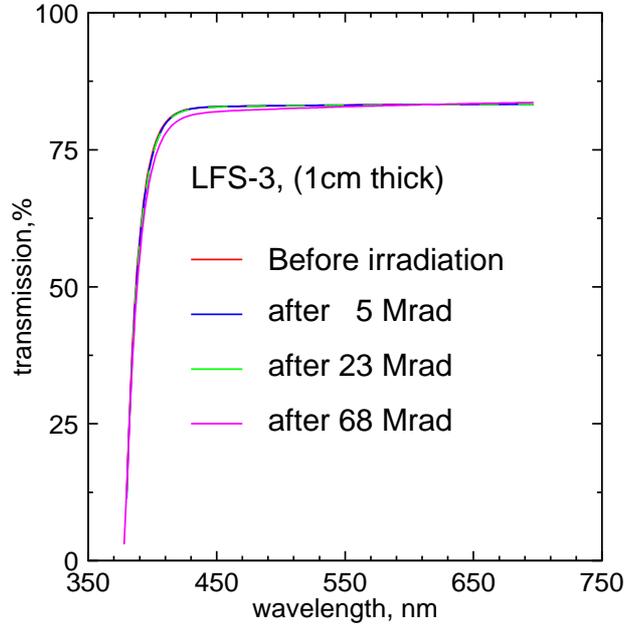}
\caption{\normalsize Transmission spectra of $LFS-3$ crystal before and after 
          irradiation(sample thickness 10 mm)} 
\label{fig:trans}
\end{figure} 

Currently there is a strong demand for ultra radiation resistant crystals
for electromagnetic calorimeters located near beam-pipe, in the endcap region,
and capable of working under heavy radiation conditions during an extended 
length of time.

In this work we study radiation hardness of small $LFS-3$ samples at accumulated doses 
from low-energy gamma-ray irradiation up to 68 Mrad.

 
The $LFS-3$ crystals were grown by Zecotek Imaging Systems Pte Ltd, Division of 
Zecotek Photonics Inc., Vancouver, Canada with the Czochralski technique. The $10\times10\times10$ $cm^{3}$ 
samples (from top, middle, bottom) were cut from $LFS-3$ boule of 10 cm diameter and 20 cm 
length then polished to an optical grade. Optical transmission spectra across a 10 mm thickness 
were measured with a spectrophotometer (Kruess Optronic VIS 6500).

Radiation hardness of $LFS-3$ samples was studied by comparing transmission spectra of 
the samples before and after irradiation. The irradiation was carried out using $^{60}Co$ source 
(maximum power is about 4 Krad/min). All $LFS-3$ crystals (top, middle, bottom of boule) were 
sequentially irradiated with three doses: 5 Mrad, 23 Mrad and 68 Mrad. Optical transmission 
spectra measurements were performed just after irradiation. The result for one $LFS-3$ crystal is 
plotted in Fig.\,\ref{fig:trans}.

From the analysis of the spectra it can be seen that with an increase of the irradiation dose 
the transmission drops down for the 68 Mrad dose only. There was no reduction in transmission 
spectra for $LFS-3$ after irradiation with the dose 23 Mrad, for samples produced from top, middle 
and bottom of a large $LFS$ boule.

\section {Conclusions}

Earlier radiation hardness of $LSO$ and $LYSO$ crystals has been already investigated 
\cite{Kobayashi,Kozma,Chen}. 
For example, the $Ce:LSO$ degradation in optical transmission after irradiation with $^{60}Co$ 
gamma-rays is about $\sim2.5\%$ per cm at 10 Mrad \cite{Kobayashi}. $LFS-3$ is a faster scintillator with better 
radiation hardness, making it a very suitable scintillation material for electromagnetic 
calorimeter used in high-energy particle physics experiments.


\end{document}